
\def\tw{\tilde}

\centerline{\bf Weaving a classical geometry with quantum threads}
\vskip1cm
\centerline{Abhay Ashtekar${}^\dagger$, Carlo Rovelli${}^\ddagger$ and Lee
Smolin${}^\dagger$}
\vskip1cm
\centerline{\it  ${}^\dagger$ Department of Physics, Syracuse University,
Syracuse, NY  13244-1130,  U.S.A.}
\vskip.5cm
\centerline{\it ${}^\ddagger$ Department of Physics, University of
        Pittsburgh, Pittsburgh, PA  15260; and}
\centerline{\it Dipartimento di Fisica, Universita di Trento, Italia}
\vskip2cm
\centerline{Abstract}
Results that illuminate the physical interpretation of states of
nonperturbative quantum gravity are obtained using the recently
introduced loop variables. It is shown
that: i) While local operators such as the metric at a point may not be
well-defined, there do exist {\it non-local} operators, such as the area of
a given 2-surface, which can be regulated diffeomorphism invariantly and
which are finite {\it without} renormalization;  ii)there exist quantum
states which approximate a given flat geometry at large scales, but
such states exhibit a discrete structure at the Planck scale.
\vskip2cm

It is by now generally accepted that perturbative approaches to quantum
gravity fail because they assume that space-time geometry can be
approximated by a smooth continuum at all scales. What is needed are
non-perturbative approaches which can {\it predict} --rather than assume--
what the true nature of the micro-structure of this geometry is. In such
an approach, background fields such as a classical metric or a
connection cannot play a fundamental role; quantum theory must be formulated
in a diffeomorphism invariant fashion. An important task in these programs
is then to introduce techniques needed to describe geometry and to
``explain'' from first principles how smooth geometries can arise on
macroscopic scales.

Over the past five years, two avenues have been pursued to test if quantum
general relativity can exist non-perturbatively: the first is based on
numerical simulations [1], while the second is based on canonical
quantization[2-6]. This letter concerns the second approach.
While the canonical approach itself was introduced by Dirac in the late
fifties, the recent work departs from the early treatment in
two important ways: i) it is based on a new canonically conjugate pair,
the configuration variable being a connection [2,5]; and, ii) it uses
a new representation in which quantum states arise as suitable functions
on the space of closed loops on a (spatial) 3-manifold [2,3,6]. The new
ingredients have led to technical as well as conceptual simplifications
which, in turn, have led to a variety of new results. In particular,
these methods have opened up new bridges between quantum gravity and
other areas in mathematics and physics such as knot theory, Chern-Simons
theory and Yang-Mills theory.

The purpose of this letter is to report on the picture of quantum geometry
that arises from the use of the loop variables. To explore geometry
non-perturbati\-ve\-ly, we must first introduce operators that carry the metric
information and regulate them in such a way that the final operators do
{\it not} depend on any background structure introduced in the regularization.
We will show that such operators do exist and that they are finite without
renormalization. (Furthermore, some of the operators are invariant under
3-dimensional diffeomorphisms and have a well-defined action on diffeomorphism
invariant states which, in the loop representation, depend only on the
knot class of the loop.) Using these operators, we seek non-perturbative
states which can approximate any given flat classical geometry up terms
${\cal O}(l_p/L)$ where $l_P$ is the Planck length and $L$ is a macroscopic
length scale, lengths being defined by the given geometry.
We find that such states do exist but that they exhibit a discrete
structure at the Planck scale $l_P$. Such a result was anticipated on general
grounds since the 30's [8]. Indeed, there exist a number of quantum gravity
programs that {\it begin} by postulating discrete structures at the Planck
scale and then attempt to recover from it the known macroscopic physics[9].
The key difference is that, in our approach, discreteness is the {\it output}
of
the framework rather than input.   The input is only general relativity
and quantum mechanics.

In this letter, we will only sketch the main ideas involved; details are
described in [3,4,7].

Let us begin with the classical phase space. The configuration
variable $A_a^i$ is a complex, $SU(2)$-connection and its conjugate momentum,
$\tw{E}^a_i$ --the mathematical analog of the electric field in Yang-Mills
theory-- is a triad with density weight one [5]. (Throughout we will let
$a,b, ...$ denote the spatial indices and $i,j,...$, the internal indices.
A tilde over a letter will denote a density weight 1.)
The first step is the introduction of loop variables [6] which are
manifestly $SU(2)$-gauge invariant functions on the phase space. The
configuration variables are the Wilson loops: Given a closed loop $\gamma$
on the 3-manifold $\Sigma$, we set
$$
        T[\gamma ]  = \textstyle{1\over 2} Tr {\cal P} \exp\,
        {G \oint_\gamma A_a dl^a}, \eqno(1)
$$
where $G$ is Newton's constant. (Throughout, we use the 2-dimensional
representation of the gauge group to
evaluate traces.) Variables with momentum dependence are constructed by
inserting $E^a_i$ at various points on the loop before taking the trace.
Thus, for example, the loop variable quadratic in momenta is given by:
$$
        T^{aa'} [\gamma ] (y,y')= Tr\left [({\cal P} \, \exp
        {G \int_{y'}^y A_a dl^a})\, \tilde{E}^a(y)\, ({\cal P}
        \exp {G \int_y^{y'} A_a dl^a})\, \tilde{E}^{a'}(y') \right ],
        \eqno(2)
$$
where $y$ and $y'$ are any two points on the loop $\gamma$. Note that
in the limit when the loop $\gamma$ shrinks to a single point $x$,
$T^{aa'}[\gamma]$ tends to $\tw{E}^a_i(x) \, \tw{E}^{a'i}(x)$, which,
when $\tw{E}^a_i$ is invertible, is related to the metric $q_{ab}$
by ${\rm det}q(x) q^{aa'}(x)=\tw{E}^a_i(x) \, \tw{E}^{a'i}(x)$.
Thus, one can recover the metric from the
loop variable $T^{aa'}$.

In quantum theory, states are represented by suitably regular
functions $\Psi [\alpha ]$ of loops satisfying certain algebraic
conditions and quantum operators corresponding to
the loop variables are defined in such a way that the Poisson algebra
of the loop variables is mirrored in the commutator algebra in the usual
fashion. For example, the action of the loop operator $\hat{T}[\gamma ]$
is given by:
$$
\hat{T}[\gamma ]\circ \Psi[\alpha ] = \Psi[\alpha \sharp \gamma ] +
                  \Psi[\alpha \sharp \gamma^{-1}], \eqno(3)
$$
where $\alpha\sharp\gamma$ is an "eye glass loop" which is equal
to $\alpha \circ \eta\circ \gamma\circ \eta^{-1} $ for an
arbitrary segment $\eta $ joining
$\alpha$ and $\gamma$ and the same segment is  used in both
terms[10]. Similarly, the action of higher order loop operators such as
$\hat{T}^{aa'}(y,y')$ simply involves gluing, breaking and re-routing of
loops.

It is tempting to try to define the local metric operator
as the limit of $\hat{T}^{ab}$. However, the resulting operator has to
be regulated and then renormalized--it
involves products of $\tw{E}^a_i$ and $\tw{E}^b_i$
evaluated at the {\it same} point-- and, because of the density weights
involved, the renormalized operator carries an imprint of the
background structure used in this procedure. This is because the
renormalization procedure changes the density weight as it must
replace a product of delta functions by a single delta function and,
on a manifold, delta functions are densities.  (In Minkowskian field
theories, there {\it is} a preferred background metric and the only
ambiguity in defining analogous operators is that of a multiplicative
renormalization {\it constant}.) This appears to be a general feature of
diffeomorphism invariant theories and it obstructs the introduction of
meaningful {\it local} operator-valued distributions carrying geometric
information.

Fortunately, however, there do exist non-local operators carrying the same
information. We now sketch the construction of two of these.

Note first that, given a smooth 1-form $\omega_a$ on $\Sigma$, we can define
a function $Q[\omega]$ on the classical phase space which carries the
metric information:
$$
 Q[\omega] := \int_\Sigma d^3x\, (\tw{E}^a_i\omega_a\,\, \tw{E}^{a'i}
  \omega_{a'})^{1\over 2} ,
\eqno(4)
$$
where the integral on the right is well-defined because the integrand is
a density of weight 1. When the triads are smooth, we can reconstruct the
density weighted inverse metric from the knowledge of $Q[\omega ]$ (for all
$\omega$). In terms of the classical loop variable $T^{aa'}$, this function
can be re-expressed as:
$$
        {Q}[\omega ] = \lim_{\epsilon \rightarrow 0}
        \int d^3x \left ( \int d^3y \int d^3y' \, f_\epsilon(x,y)
        f_\epsilon(x,y') T^{aa'}[\gamma_{y,y'}](y,y')
\omega_a (y) \omega_{a'}(y')
        \right )^{1 \over 2}, \eqno(5)
$$
where $f_\epsilon(x,y)$ is a smearing function, a density of weight 1 in
$x$, which tends to $\delta^3(x,y)$ as $\epsilon$ tends to zero and where
$\gamma_{y,y'} $ is an arbitrarily defined
smooth loop that passes through points
$y$ and $y'$, such that it goes smoothly to a point as
as $y' \rightarrow y$ .
Expression (5) of $Q[\omega ]$ is well-suited for translation
to quantum theory: We can define the quantum operator $\hat{Q}[\omega ]$
simply by replacing $T^{aa'}$ in (5) by the loop
operator $\hat{T}^{aa'}$ and taking the limit $\epsilon \rightarrow 0$
in the action of the operator on states.
The resulting operator is well-defined -- it carries no memory of the
additional structure used in the construction of the smearing functions--
and finite without any renormalization [11].  The resulting action of the
operator on states is quite simple, if $\alpha$ is a nonintersecting[12]
loop it is
$$
   \hat{Q}[\omega ]\circ \Psi [\alpha ] = (\sqrt{6}\, l_P^2\,
   \oint_\alpha ds
   |\dot{\alpha}^a \omega_a(\alpha(s))|)\,\, \Psi(\alpha ).\eqno(6)
$$
Thus, the operator acts simply by multiplication. Hence the loop
representation is well-suited to find states in which the 3-geometry
--rather than its time evolution-- is sharp.

The second class of operators corresponds to the area of 2-surfaces.
Note first that, given a smooth 2-surface S in $\Sigma$, its area
${\cal A}_S$ is a function on the classical phase space. We first express
it using the classical loop variables. Let us divide the surface $S$ into
a large number $N$ of area elements $S_I, I=1,2...N$, and set
${\cal A}_I^{\rm appr}$ to be
$$
  {\cal A}_I^{\rm appr} =\left[ \int_{S_I} d^2S^{bc}(x) \epsilon_{abc}
   \int_{{\cal S}_I} d^2S^{\prime b'c'} (x') \epsilon_{a'b'c'}\,
   T^{aa'}[\gamma_{x,x'} ](x,x') \right]^{1\over 2}, \eqno(7)
$$
where $\epsilon_{abc}$ is the (metric independent) Levi-Civita density of
weight -1.   ${\cal A}_I^{\rm appr}$ approximates
the area function on the phase space defined by the surface elements $S_I$,
the approximation becoming better as $S_I$ --and hence
loops $\gamma_{x,x'}$--
shrink. Therefore, the total area ${\cal A}_S$ associated with $S$ is given
by
$$
   {\cal A}_S = \lim_{N \rightarrow\infty}\, \, {\cal A}_I^{\rm appr}.
   \eqno(8)
$$
To obtain the quantum operator $\hat{\cal A}_S$, we simply replace
$T^{aa'}$ by the quantum loop operator $\hat{T}^{aa'}$. This somewhat indirect
procedure is necessary because there is no well-defined operator that
represents the metric or its area element {\it at a point}. Again, the
operator $\hat{\cal A}_S$ is finite and its action is  simple when
evaluated on a nonintersecting[12] loop $\alpha$:
$$
  \hat{\cal A}_S \circ \Psi[\alpha] =
   \sqrt{6}\, l_p^2 \, I(S,\alpha )\, \Psi[\alpha],\eqno(9)
$$
where $I(S,\alpha)$ is simply the unoriented intersection number between
the 2-surface $S$ and the loop $\alpha$ [4]. Thus, in essence, a loop
$\alpha$ contribute one Planck unit of area to any surface it intersects.
The area  operator also acts simply by multiplication in the loop
representation.

Because of the simple form of operators $\hat{Q}[\omega ]$ and
$\hat{\cal A}_S$, a large set of simultaneous eigenstates
can be immediately constructed.
There is one associated to every nonintersecting loop $\gamma$, which
we will label $\Psi_\gamma [\alpha ]$ and call the characteristic
state of $\gamma$. It is equal to one when
evaluated on $\gamma$ and zero when $\alpha $ is
any other nonintersecting loop[13].  Its value on intersecting loops
may be found by an explicit computation, which is
described in refs.[4,7].  We may note that
the corresponding eigenvalues of area are then {\it quantized}
in integer multiples of $\sqrt{6}l_P^2$.  There are also
eigenstates associated with intersecting loops; these are discussed
in refs.  [4,7].

Let us now turn to the second of our main results. The goal here is to
introduce loop states which approximate a given flat 3-metric $h_{ab}$
on $\Sigma$ on scales $L$ large compared to $l_p$. (Note that the
large scale limit is equivalent to the semi-classical limit since, in
source-free, non-perturbative quantum general relativity, $\hbar$ and
$G$ always occur in the combination $\hbar G = l_p^2$.) The basic idea
is to weave the classical metric out of quantum loops by spacing them so
that (on an average) precisely one line crosses every surface element
whose area, {\it as measured by the given} $h_{ab}$, is one Planck unit.
Such loop states will be called {\it weaves}. Given a weave, one can
obtain others by, e.g., adding small quantum fluctuations. We now
present a concrete example of such a state.

Using the given flat metric $h_{ab}$, fix a cubical lattice on $\Sigma$
with lattice spacing $a$. At each lattice site $\vec{n}$ we center a
circle $\gamma_{\vec{n}}$ of radius $a$ and random orientation (where
$h_{ab}$ is again used in the construction, and we require also
that the loops be, for simplicity, nonintersecting.) Denote the collection
of these circles by $\Delta$. The state $\Psi_\Delta (\alpha )$ is, as we now
indicate, a weave state with the required properties. To see if it
reproduces on a scale $L>>l_p$ the geometry determined by the classical
metric $h_{ab}$, let us introduce a 1-form $\omega_a$ which is slowly
varying on the scale $L$ and compare the value $Q[\omega](h)$ of the classical
$Q[\omega]$ evaluated at the metric $h_{ab}$, with the action of the quantum
operator $\hat{Q}[\omega]$ on $\Psi_\Delta [\alpha ]$. A detailed calculation
yields:
$$
\hat{Q}[\omega]\circ\Psi_\Delta[\alpha] =  \left[ 2\pi\sqrt{6}\,
({l_p\over a})^2 \, Q[w](h) + {\cal O}({a\over L})\right]\,
\Psi_\Delta [\alpha ].\eqno(10)
$$
Thus, $\Psi_\Delta$ is an eigenstate of
$\hat{Q}[\omega ]$ and the corresponding eigenvalue is closely related to
$Q[\omega](h)$. However, even to the leading order, the two are unequal
{\it unless} the lattice separation $a$ {\it equals} $(2\pi\sqrt{6})^{1/2}\,
l_P$.

The situation is the same for the area operators $\hat{\cal A}_S$. Let
$S$ be a 2-surface whose extrinsic curvature varies slowly on a scale
$L >>l_P$. The state $\Psi_\Delta$ is an eigenvector of $\hat{\cal A}_S$
with eigenvalue equal to the area ${\cal A}_S(h)$ assigned to $S$ by
$h_{ab}$ (to ${\cal O}(l_P^2/{\cal A}_S(h)$) when the lattice spacing
$a$ satisfies precisely the condition stated above.

Thus, the requirement that $\Psi_\Delta$ should approximate the classical
metric $h_{ab}$ on {\it large scales} $L$ tells us something non-trivial
about the {\it short-distance structure} of the multi-loop $\Delta$: $a$
is fixed to be the Planck length in units defined by $h_{ab}$. Now,
naively, one might  have expected that the best approximation to the classical
metric would occur in the continuum limit in which the lattice separation
goes to zero. It is, perhaps,
surprising that this does not occur. The reason
is that the factors of the Planck length
in (6) and (9) force each line of the weave to contribute a Planck
unit to the various geometrical observables.  Finally,
note the structure of the argument: we begin with a classical metric,
use it to define the scale $L$, the notion of ``slowly varying''  as well as
the structure of $\Delta$ and find that the lattice separation $a$ is
forced to be the Planck length as measured by $h_{ab}$. This is
necessary because,  while
the theory depends on a dimensional scale $l_p$, it is only with
respect to those quantum states that can be associated with classical
metrics in this way that there can be a
meaningful specification of what physical intervals have this
length.

We conclude with three remarks.

1) Work is in progress on the linearization of the exact state space of
the theory  in a neighborhood of a weave state. There are preliminary
indications [3,14] that the resulting theory is isomorphic to the Fock space
of linearized gravitons[15] to order $l_{P}/ \lambda_{\rm graviton}$.
Thus, the notion of gravitons may be physically meaningful
only at long wave
lengths.  In this connection it is important to point out that
as it is an eigenstate of the three geometry $\Psi_\Delta [\alpha ]$
is not a candidate for the vacuum of the theory; however candidates
for the vacuum may be constructed by dressing $\Psi_\Delta [\alpha ]$
with an appropriate distribution of loops corresponding to the
virtual gravitons.\hfil\break
2) The main results presented in this letter can be obtained also in the
connection representation [2,5,16], in which case the
characteristic states take the form
$\Psi_\gamma [A] = Tr {\cal P} exp( G \oint_\gamma A_ad\gamma^a)$.
It is the loop operators (rather than
the loop states) that are essential to the argument; they provide us with
a regularization procedure that respects diffeomorphism invariance.
\hfil\break
3) However, if we wish to consider physical states which are annihilated by
the quantum constraints the use of the loop representation seems
unavoidable since, at the present stage, solutions to all quantum
constraints have been obtained only in the loop picture. Consider the
loop state $\Psi_{\rm flat}$ defined on nonintersecting loops $\alpha$ by:
$$
     \Psi_{\rm flat} [\alpha] = \cases
            {1 & \hbox{if $K(\alpha)=K(\Delta)$ } \cr
             0  & \hbox{otherwise,}\cr}\eqno(11)
$$
where $K(\alpha)$ is the knot class to which $\alpha$ belongs.
Results of [6] show that $\Psi_{\rm flat}(\alpha )$ solves all
quantum constraints; in the Dirac terminology, it is a {\it physical state}
of non-pertubative quantum gravity. It is natural to interpret
$\Psi_{\rm flat}$ as representing the flat 3-geometry at large scales:
just as the loop $\Delta$ corresponds to a flat metric $h_{ab}$, the
equivalence class $K(\Delta)$ of loops should correspond to the the
equivalence class of all metrics related to $h_{ab}$ by a diffeomorphism.
In the same spirit then, it is tempting to conjecture that knot classes
which are inequivalent to $K(\Delta)$ represent non-flat geometries.

\bigskip
{\bf Acknowledgments:} This work was supported in part by the NSF
grants PHY90-12099, PHY90-16733 and INT88-15209 and by research funds
provided by Syracuse University.
\bigskip

\item{[1]} B.S. DeWitt, E. Myers, R. Harrington and A. Kapulkin,
Nucl. Phys.B. (Proc. Suppl.) {\bf 20}, 744 (1991); M.E. Agishtein
and A. A. Migdal, Princeton pre-print PUPT-1272 (1991).

\item{[2]} A. Ashtekar, {\it Non-perturbative canonical quantum gravity}
(Notes prepared in collaboration with R.Tate) (World Scientific, Singapore,
1991).

\item{[3]}C. Rovelli, Class. \& Quant. Grav. {\bf 8}, 1613 (1991).

\item{[4]} L. Smolin,  {\it Recent developments in
nonperturbative quantum gravity}  Syracuse pre-print SU-GP-92/2-2,
to appear in the Proceedings of the 1991 GIFT International Seminar
on Theoretical Physics (World Scientific, Singapore, in press.), hepth
preprint number 9202022.

\item{[5]} A. Ashtekar Phys. Rev. Lett. {\bf 57}, 2244 (1986),
Phys. Rev.{\bf D36}, 1587 (1987).

\item{[6]} C. Rovelli and  L. Smolin, Phys. Rev. Lett. {\bf 61}, 1155
(1988); Nucl. Phys. {\bf B133}, 80 (1990).

\item{[7]}A. Ashtekar, C. Rovelli and L. Smolin, in preparation.

\item{[8]}  See, for example, J. A. Wheeler, in "Battelle Rencontres
1967", eds. C. DeWitt, J. A. Wheeler,
Benjamin 1968; B. Ferretti, Lettere al Nuovo Cimento, 40, 169 (1984).;
M. Toller, International Journal of Theoretical Physics, 29, 963 (1990);
T. Padmanbhan, Class. and Quantum Grav. 4, L107, (1987).

\item{[9]}See, for example, D. Finkelstein, Phys. Rev. 184 (1969) 1261;
 R. Penrose, in {\it Quantum Theory and Beyond},
ed. T. Bastin (Cambridge University Press,1971); G. 't Hooft,
{\it Recent Developments in Gravitation}, eds. in M. Levy
and S. Deser, (Plenum,1979); L. Bombelli, J. Lee, D. Meyer and R. D.
Sorkin, Phys. Rev. Lett. 60 (1988) 655; J. Barbour and L. Smolin,
Syracuse Preprint SU-GP-92/2-4 (1992).

\item{[10]}This definition is equivalent to the one given in [2-4,6]
in which the state space is extended to functions on multiloops
and $\hat{T}[\gamma ]\circ \Psi[\alpha ] = \Psi[\alpha \cup \gamma ]$.

\item{[11]}  In fact the finiteness of these operators is closely related
to their background independence. Basically, a dependence on a power of the
regulator must come together with background dependence, because the
regulator scale is defined with respect to the background metric
introduced in regularization. It is tempting to conjecture that
any operator that is regulated in such a way that no dependence on
the background survives in the limit in which the regulator is
removed, must be finite in this limit.

\item{[12]} The action of the operator on intersecting loops is known
explicitely, but is slightly more complicated.  See refs. [4] and [7] for
details.

\item{[13]}Note that such states are not normally considered in
ordinary quantum field theories based on Fock inner products.  However,
in the absence of a background metric the Fock inner product cannot
be defined.  Instead, quantum field theories on manifolds without
background metrics may be based on representations of the sort
introduced by A. Ashtekar and C.J. Isham (Class. \& Quan. Grav., in
press) within which the characteristic states of loops
{\it are} normalizable.

\item{[14]}J. Zegwaard, Utrecht preprint THU-91/13 (1991).

\item{[15]}A Ashtekar, C. Rovelli and L. Smolin, ``Gravitons and loops''
Phys. Rev. D  44 (1991) 1740.

\item{[16]} T. Jacobson and L. Smolin, Nucl.Phys. {\bf B299}, 295 (1988).

\end